# Crossover behavior in the distance dependence of hydrophobic force law

Tuhin Samanta, Saikat Banerjee and Biman Bagchi*

Solid State and Structural Chemistry Unit, Indian Institute of Science, Bangalore – 560012, India

E-mail: bbagchi@sscu.iisc.ernet.in

## Abstract

**Understanding about both the range and the strength of the effective force between two hydrophobic surfaces suspended in water is important in many areas of natural science but unfortunately has remained imperfect. Even the experimental observations have not been explained quantitatively. Here we find by varying distance (d) between two hydrophobic walls in computer simulations of water that the force exhibits a bi-exponential distance dependence. The long range part of the force can be fitted to an exponential force law with correlation length of 2 nm while the short range part displays a correlation length of only 0.5 nm. The crossover from shorter range to longer range force law is rather sharp. We show that the distance dependence of the tetrahedrality order parameter provides a reliable marker of the force law, and exhibits similar distance dependence.**

## I. Introduction

Origin and magnitude of an attractive hydrophobic interaction[1-6] among non-polar solute species or surfaces suspended in water have been studied extensively because of its central role in chemistry and biology. In 1982, Israelachvili[7-8] first discovered that an attractive force can exist between two hydrophobic walls immersed in water. This attractive force was found to be surprisingly long ranged (e.g. 10 – 100nm) and decayed exponentially with distance. This became widely known as the hydrophobic force law (HFL). An even longer range force that was found was later attributed to surface fluctuations or modulations. An elegant theory of HFL was developed by Lum, Chandler and Weeks (LCW)[9] that explained the origin of the hydrophobic force detected by Israelachvili[7-8], in terms of density functional theory where free energy of the system was expanded in terms of density fluctuations. However, many aspects of hydrophobic force law (HFL)[10-11] have remained ill-understood. In particular it was originally suggested that the long ranged order of interaction implied an electrostatic coupling between the two surfaces. Subsequently more refined experiments ruled out electrostatic origin because the force showed only a weak dependence on added electrolyte. At small separation $d$, simulations have observed a drying transition that was also first predicted by LCW. This phenomenon has been attributed, at least in part, to a theoretically predicted cavitation at some critical inter surface separation. Theoretical studies have shown that the degree of hydrophobicity, size of the hydrophobic surface and temperature or pressure dominantly influence the critical separation as well as the rate of evaporation of the confined water[12-13]. Such cavitation (or dewetting transition) was earlier observed near liquid-vapor coexistence for Lennard-Jones fluid confined between hydrophobic walls[14].

Recently several experiments[15-17] have been performed which show that any intrinsic hydrophobic force originating from orientation of water molecules at hydrophobic surfaces is quite short ranged with a decay length of only 3 to 4 angstroms. Interestingly, an exponential correlation length of 3.8 angstrom is also predicted by the Lum, Chandler and Weeks theory[9, 18] of water structure at air-water interface.

Water under ambient conditions is a locally orientationally ordered liquid, with the local orientational order parameters not too much lower than those in ice. In liquid water, however, this local orientational order fails to propagate beyond a couple of nearest neighbor separations. Our previous study[19] has shown that Mercedes Benz (MB) water model qualitatively reproduces many properties of water, also reproduces the hydrophobic force law. It has been found that orientation of the molecules near and between the hydrophobic surfaces play important role at longer length scale. This model has also been used to study the thermodynamic properties and structural aspects of confined water. We have found that a destructive interference between orientation heterogeneity propagating inwards from the two surfaces reduces orientational heterogeneity thus lowering free energy. Therefore the effective attraction between two surfaces increases as the destructive interference increases with decreasing separation between the two walls.

Confinement of water between two surfaces can show a significantly different behavior from those observed in the bulk. Here we study the role of orientation of the water molecules confined between two hydrophobic walls.

## II. Description of the model and simulation details

Simulations are performed using Large scale Atomic/Molecular Massively Parallel Simulator (LAMMPS) MD packages. In this study, all the molecular dynamics simulations are carried out using extended single point charge water (SPC/E) model[20] in a cubic box. We have mimicked the arrangements of carbon atoms in grapheme sheets. The hydrophobic walls (10 X 10 angstrom$^2$ and 20 X 20 angstrom$^2$) are represented by a rigid, hexagonal, lattice of Lennard-Jones (LJ) atoms with a lattice constant 1.4 angstrom. The LJ particles are considered to be mutually non-interacting. The walls are kept fixed and are separated by a distance of *d* (see **Figure 6**). We have inserted the two walls at required distance by replacing any water particles, if within contact distance. Further details can be found in the Appendix.

## III. Results and discussion

### A. Hydrophobic force law

The pressure of the region confined between the two hydrophobic walls is obtained from the virial expression,

$$P_{cav} = \frac{Nk_B T}{V} + \frac{\sum_i r_i f_i}{dV} \tag{1}$$

where $N$ is the number of atoms, $k_B$ is the Boltzmann constant, $T$ is the temperature, $d$ is the dimensionality of the system, $V$ is the volume and the second term of the **Eq.1** is the virial.

In **Figure 1(a)**, we show the force on the walls as a function of $d$ (distance between two hydrophobic walls) for a particular dimension ($A$) of the walls. The net force on the walls can be obtained by multiplying the effective pressure with the respective area of the walls, and should also involve directionality. The negative force indicates the attraction between the walls induced by the water molecules. The two walls are considered to be non-interacting among them.

The effective force ($F$) on the walls is obtained as, $F = F_{cav} - F_{\infty}$. We obtain the force at infinite separation ($F_{\infty}$) between the two walls by fitting. This effective force is now given by the following equation,

$$F = a_{F_1} \exp(-d/\xi_{F_1}) + a_{F_2} \exp(-d/\xi_{F_2}). \tag{2}$$

where, $d$ is the inter-plate distance, $a_{F_1}$, $a_{F_2}$, $\xi_{F_1}$ and $\xi_{F_2}$ are fitting parameters. In the long range limit this is consistent with the experiments of Israelachvili *et. al.*[7] As shown in **Figure 1(a)**, the fitting to the bi-exponential form is satisfactory, with the fitting parameters given by $a_{F_1} = -1.39$, $a_{F_2} = -0.04$, $\xi_{F_1} = 4.93$ and $\xi_{F_2} = 19.40$.

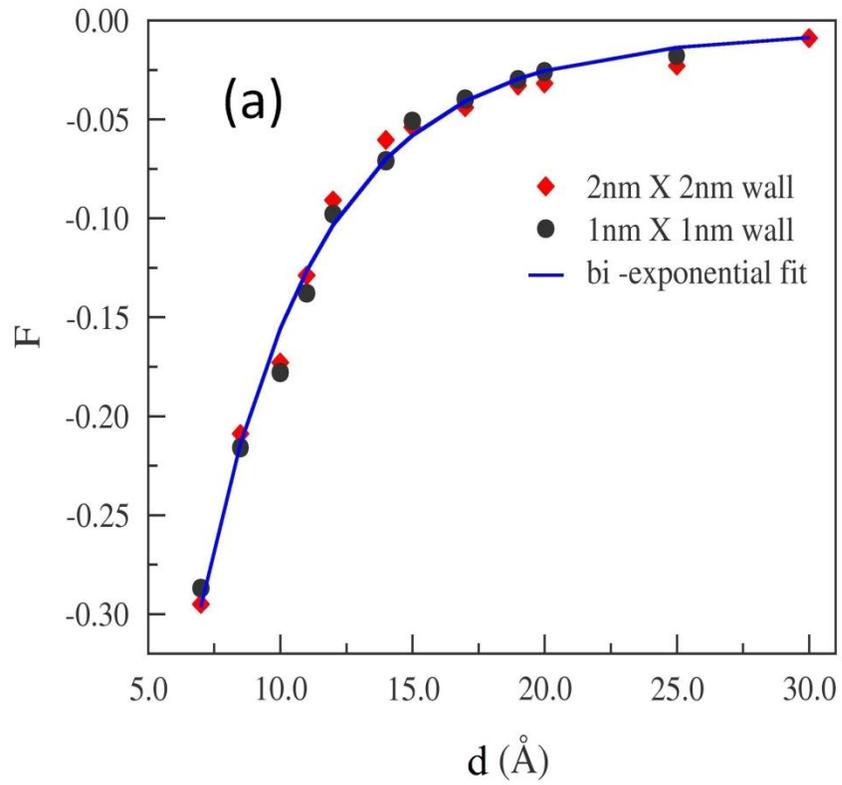

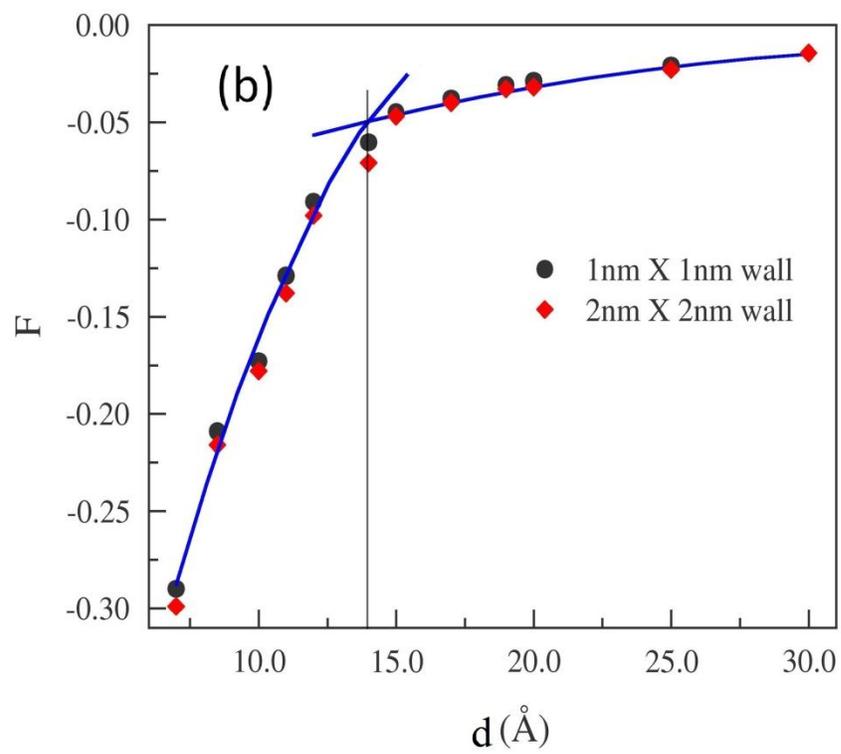

**Figure 1: (a) Force (F) on the hydrophobic walls suspended in SPC/E water increases exponentially as the distance between two walls gradually increases. The solid blue line is the bi-exponential fit with correlation lengths of 4.93 and 19.40. (b) A clear crossover is observed in the hydrophobic force law near 14 angstrom separation between two hydrophobic walls. The solid blue lines are the exponential fit.**

We find the correlation lengths, $\xi_i$, are nearly independent of the dimension of plate. The global fitting parameters or the correlation lengths are 4.93 angstrom and 19.40 angstrom. The hydrophobic force law is well demonstrated by our calculation.

**Figure 1(b)** demonstrates a sharp crossover in the attractive force law that occurs near $d = 14$ angstrom. This resembles the drying transition predicted by the Lum-Chandler-Weeks (LCW). We find one more crossover is possible below 5 angstrom (not shown) where there is hardly any volume for the water molecules expect single file movements. Experiments[15-18] have observed drying transition near 4 angstrom which is expected on physical grounds (water molecules may be forced out by direct attraction between two walls). In fact, if we plot the short range part to single exponential as shown in **Figure 1(b)** we find a correlation length of similar magnitude.

### B. Microscopic origin of the hydrophobic force law

In order to interpret the microscopic origin of the hydrophobic force, we thoroughly study two underlying order parameters – position dependent density and local tetrahedral order parameter of water molecules (confined water) located in between the two hydrophobic walls. These studies may provide new insight into the origin of the hydrophobic force.

### (i) Density profile

Theoretical approaches to understand the hydrophobic force law are based on density dependent theories, such as Lum-Chandler-Weeks (LCW) theory[9]. For a complex liquid like water, hydrogen bonding between water molecules plays an important role. Thus single order parameter descriptions might not be enough to describe the whole scenario. In particular, metastability plays a significant role in the rich behavior displayed by the confined water between two hydrophobic surfaces. Thus, it is really essential to study the density profile of water in between the two hydrophobic walls at different inter wall separations, and also at different thermodynamic conditions to understand the scope of the existing density dependent theories (such as DFT). Clearly, the water molecules between two hydrophobic walls are in dynamic equilibrium with the bulk. The distance-dependent density in between two hydrophobic walls can be determined by the standard formula

$$\rho_{cav} = \frac{\overline{N}}{V} \qquad (3)$$

Volume is defined as, $V = dA$, where, A is the wall area. $\overline{N}$ is the average number of the particles within the box of width $d$ and area $A$.

We also define a distance (z) dependent density, $\rho_{cav}(z)$, by following a coarse graining where we divided the whole cavity box into small boxes. The centre of the box serves us to give the distance z from the left wall. By symmetry, the boxes at same z should have the same local density. We therefore improved statistics by averaging over the small boxes at the same z value. The results of such calculations of local density are shown in **Figure 2.**

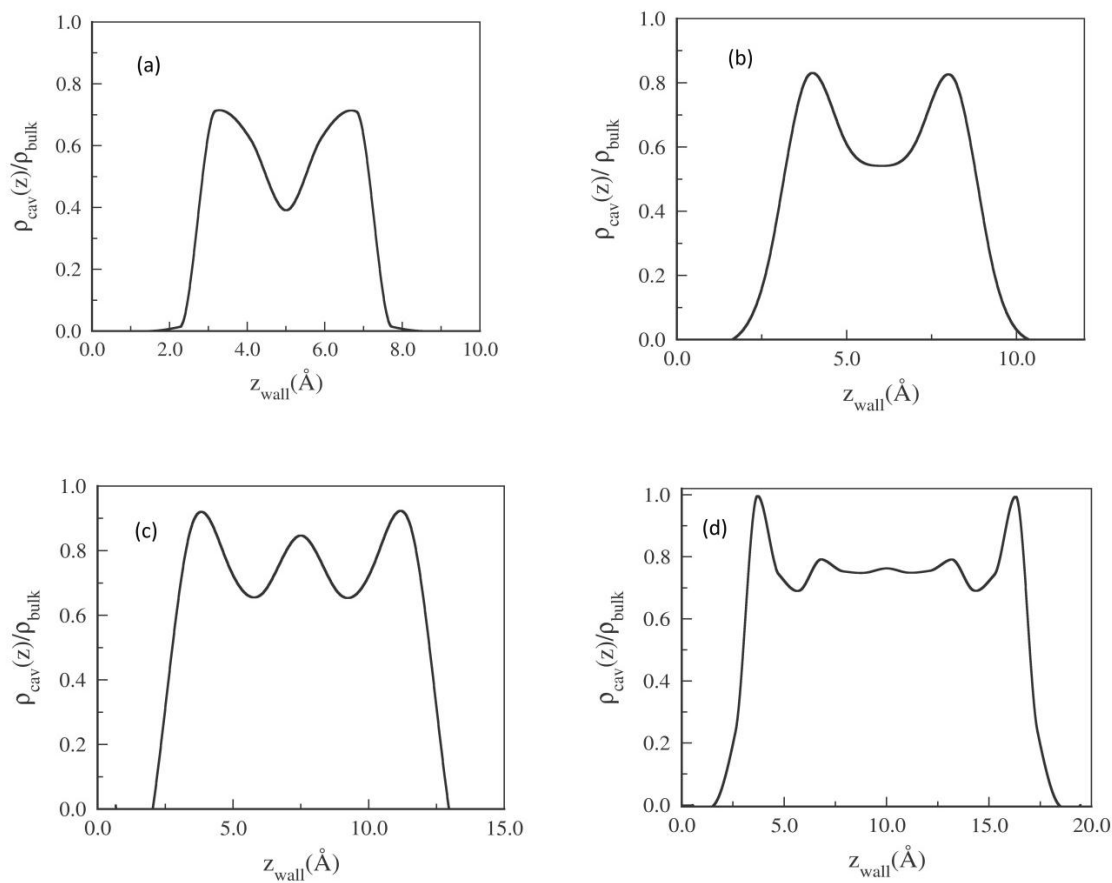

**Figure 2:** Density profile ($\rho_{cav}(z)/\rho_{bulk}$) of water between two hydrophobic walls at different inter-wall separations *d,* (a) *d* = 10 angstrom, (b) *d* =12 angstrom (c) *d* =15 angstrom and (d) *d* =20 angstrom. The local cavity density is nomalized by the bulk density of the water molecules( i.e. 0.99 g/cm$^3$). As described in the text, the local density is obtained by constructing small boxes centered around z.

Densities in **Figure 2** are normalized by bulk density. This figure shows computed water density profile at different inter wall separations, that is for different values of d. Similar profiles were also obtained by Rossky *et. al*[21-22].

Density profile oscillates along the z-direction i.e. the distance between two hydrophobic walls. As multiple layers structure is formed by water molecules between the two hydrophobic walls hence oscillation is observed in the density profile. These results are also consistent with previous work, studied by Chaudhury *et.al*[23]

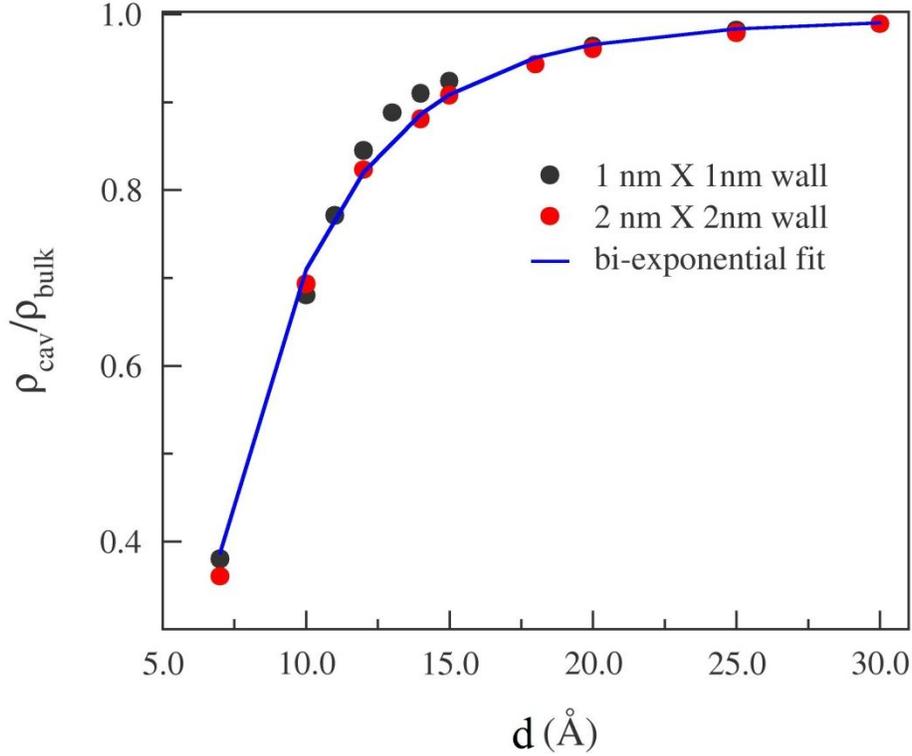

**Figure 3: Density of water in a region between two parallel hydrophobic walls, as a function of distance between two hydrophobic walls.**

In **Figure 3** we plot the average cavity densities as a function of inter wall separation, d. The solid circles (black and red) are the simulation results. The variation of the normalized cavity density is found to be bi-exponential and can be fitted to the following form,

$$\rho_{cav(d)} / \rho_{bulk} = 1 + a_{\rho_1} \exp(-d/\xi_{\rho_1}) + a_{\rho_2} \exp(-d/\xi_{\rho_2}) \tag{4}$$

where, the fitting parameters are $a_{\rho_1}$ = -3.69, $a_{\rho_2}$ = -0.060, $\xi_{\rho_1}$ = 3.86 and $\xi_{\rho_2}$ = 15.2.

Thus, we obtain the scalar density order parameters or the correlation lengths to be 3.86 angstrom and 15.2 angstrom, similar to those obtained for the force law. However, in the long range limit the density correlation is found to be somewhat less than the force correlation length. This signifies that attractive forces between two walls still persist even if the density of confined water reaches the bulk water density. Due to the shorter length scale of the density order parameter compared to the force, the microscopic origin of the attractive force between two hydrophobic walls may not be accurately described by the scalar density order parameter alone.

**(ii)        Variation of local tetrahedral order parameter with distance**

As already mentioned, for a complex liquid like water one order parameter (density dependent order parameter) description may not be enough, particularly for the confined water between two hydrophobic walls. Water molecules form the H-bond networks and this network can get compromised by the presence of the walls whose disturbing influence can propagate inside. We note that here the density of water molecules is determined to a large extent by H-bonding network.

In order to explore the effect of relative angular distribution (orientation) of water molecules in the origin of the hydrophobicity, we consider a different (but well-known) order parameter, namely the local tetrahedral order parameter which is described as,

$$t_h \equiv 1 - \frac{3}{8} \sum_{j=1}^{3} \sum_{k=j+1}^{4} \left( \cos\psi_{jk} + \frac{1}{3} \right)^2 \tag{5}$$

For, perfect tetrahedra $t_h$ reaches its maximum value unity and for a non-interacting system the average value of $t_h$ is zero. This parameter can be used to analyze the local structure or packing.

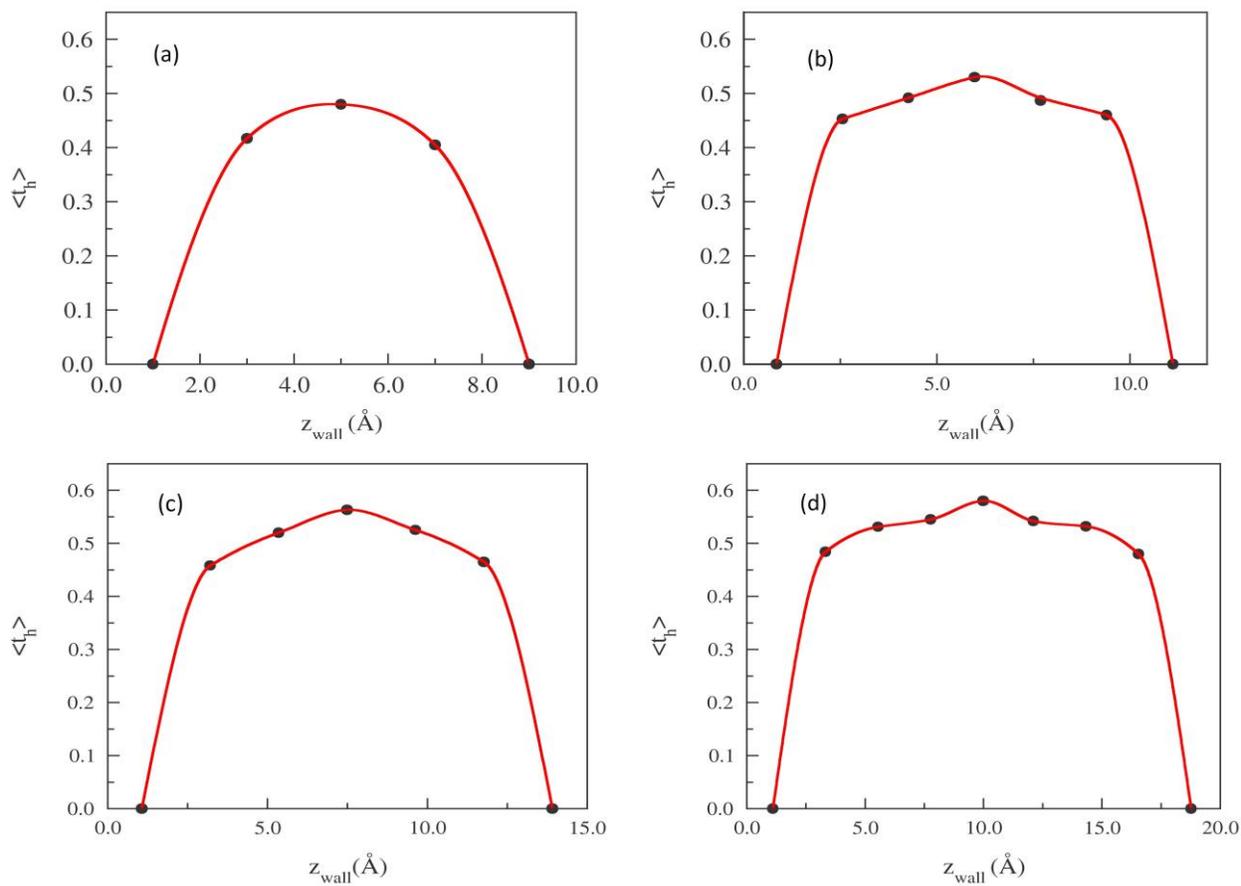

**Figure 4: The average tetrahedral order parameter value within a grid (i.e. rectangular box) along the axis of separation of the distance between two hydrophobic walls, at different inter-wall separations $d$, (a) $d$ = 10 angstrom, (b) $d$ =12 angstrom (c) $d$ =15 angstrom and (d) $d$ =20 angstrom.**

In **Figure 4** we plot the average local tetrahedral order parameter of the confined water molecules which are located in different slabs. Increasing value of this local tetrahedral order parameter indicates the structure becomes more bulk water –like. Just like in the case of the calculation of average local density, we calculate the local tetrahedrality parameter by averaging over blocks (of width 0.04 nm) located at fixed value of distance z from the left wall, at different inter plate separations, d.

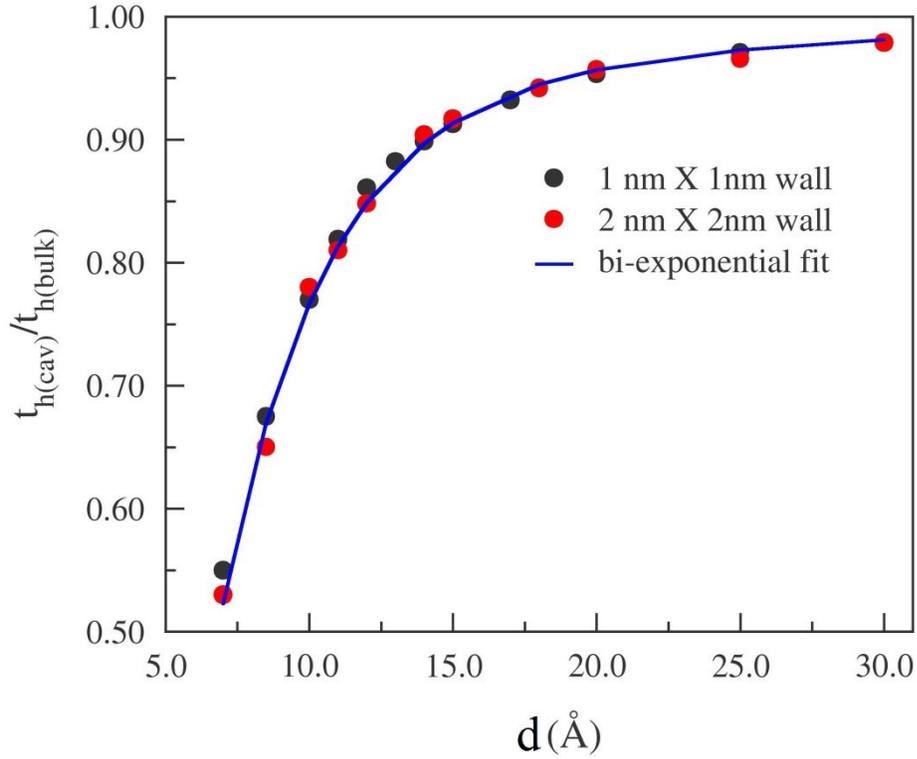

**Figure 5: Average tetrahedral order parameter ($t_h$) value of water in a region between two parallel hydrophobic walls, as a function of distance between the walls.**

The local tetrahedral order parameter value is normalized with the bulk value of the order parameter. In **Figure 5**, solid circles (black and red) are the results of simulation. The solid blue line is the bi-exponential fit of the form,

$$t_{h(d)}/t_{h(bulk)} = 1.0 + a_{t_1} \exp(-d/\xi_{t_1}) + a_{t_2} \exp(-d/\xi_{t_2}) \tag{6}$$

where, the fitting parameters are, $a_{t_1}$ = -2.96, $a_{t_2}$ = -0.095, $\xi_{t_1}$ = 3.72 and $\xi_{t_2}$ = 16.7.

Note that in the long range limit the correlation length of the local tetrahedral order parameter is a bit more comparable to the correlation length obtained from the hydrophobic force law.

## IV.     Conclusion

The main result of this work is given in **Figures 1(a)** and **1(b)**. These two figures serve to demonstrate that the attractive force law between two hydrophobic surfaces is indeed display two regimes, one short range with correlation length of about 4-5 Angstrom and a longer range with correlation length about 18-20 Angstrom. **Figure 1(b)** shows that there is a sharp cross-over between the two regimes.

While the initial landmark experiments of Israelachvili[7] indicated a rather long ranged attractive hydrophobic force law. Later experiments seem to suggest somewhat shorter range force. Clearly, the range of the force shall depend strongly on the thermodynamic conditions, and may increase in range as we approach the critical temperature and critical density. However, the magnitude of the force can also decrease at the same time.

We have shown that the attractive force correlate with the profile of the number density and the orientational order parameter. This is expected because free energy of the system varies with these parameters.

Previously we have studied MB model that has properly reproduced the hydrophobic force law[19]. The specific attention to the angular distribution/orientation of the water molecules inside the two hydrophobic walls has not been properly investigated in earlier studies. In the present study we provide a microscopic analysis of the hydrophobic force law in 3D water models. These analyses reveal that in the long range limit local tetrahedral order parameter correlation length is slightly longer than the density correlation length but both are comparable to the correlation length that we have found from the hydrophobic force law. The reported low value of correlation length in short range limit is essentially due to the intrinsic disorder of the system at low density that rapidly lowers the value of the order (density and tetrahedrality) parameters.

# Appendix

## Simulation details

We have performed the MD simulations in the isothermal-isobaric ensemble (NPT) at 298 K and 1 bar in the periodic boundary simulation box, using the Nose-Hoover[24-25] thermostat and barostat. 5000 water molecules are taken for this simulation. The system is then equilibrated for $10^5$ steps at constant temperature and volume, with each time step $\tau = 2$ fs. The production run is carried out for $10^7$ steps at constant temperature and volume. A schematic illustration of the system is given in Figure 6. The simulated system is of course three dimensional.

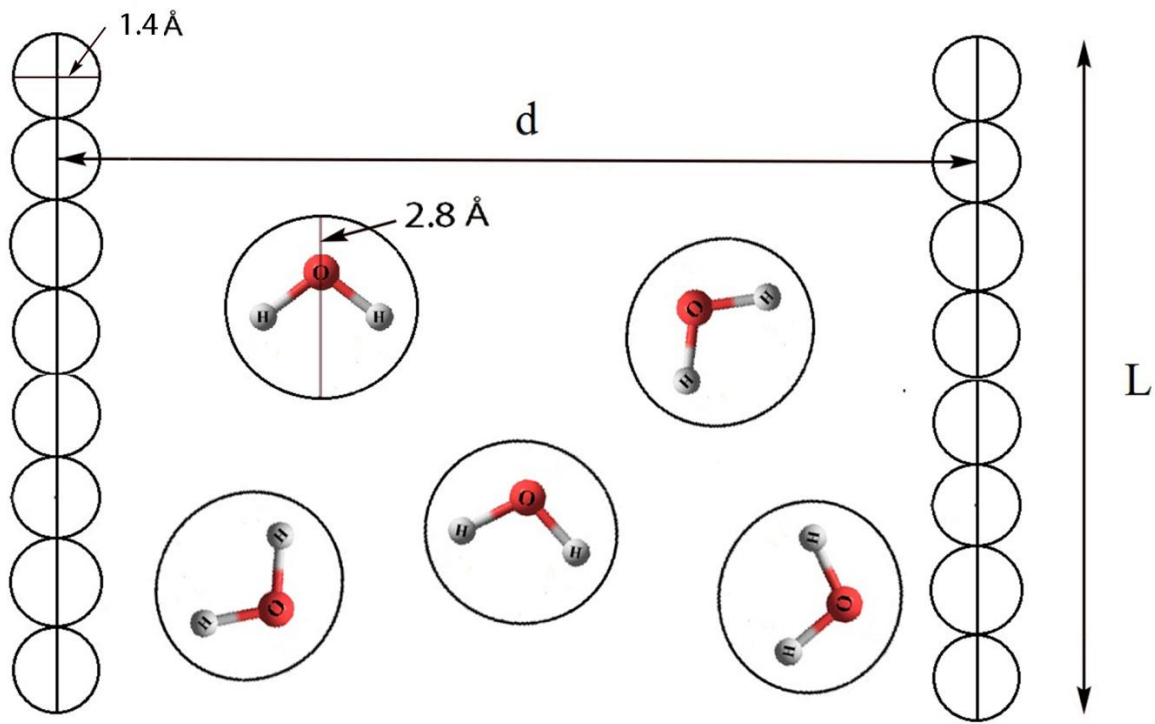

**Figure 6: Schematic representation showing the region of the hydrophobic confinement at inter-wall separation *d*. Here, *L* represents the dimension of the wall.**

The particle –particle particle –Mesh (PPPM) Ewald method is used to compute long range corrections of electrostatic interactions. The k-space is taken to be 0.0001 angstrom $^{-1}$, and calculations are performed on a 15×15×15 grid, with rms precision of $4 \times 10^{-5}$, which are the standard PPPM Ewald parameters in LAMMPS.

## Acknowledgment

It is a pleasure to thank Mr. Milan Hazra, Mr. Rajesh Dutta, Mr. Saumyak Mukherjee and Mr. Sayantan Mondal for useful discussions. BB thanks J.C. Bose Fellowship for support of the work. We also thank DST (India) for partial support.